# Multi-objective computational design optimization of a Total Disc Replacement implant

Lucia Kölle[1*], Victor Oei[1,2], Ida Mónus[1], Syn Schmitt[2], Daniel Haschtmann[3], Fabio Galbusera[4], Stephen J. Ferguson[1], Benedikt Helgason[1]

[1] Institute for Biomechanics, ETH Zürich, Zürich, Switzerland
[2] Institute for Modelling and Simulation of Biomechanical Systems, University of Stuttgart, Stuttgart, Germany
[3] Spine Surgery, Schulthess Klinik, Zürich, Switzerland
[4] Research Group Spine, Schulthess Klinik, Zürich, Switzerland

**Abstract**

While cervical arthroplasty using Total Disc Replacement (TDR) implants is an established treatment for persistent neck and arm pain, revision rates limit it from reaching its full potential. To address the underlying complications, we developed finite element simulation-driven design optimizations for a TDR's bone-implant interface and motion-preservation features. These automated processes explored high-dimensional design spaces iteratively through analysis of design variations interplay with spinal structures. The optimizations were metamodel-based using artificial neural networks and a hybrid optimizer. They optimized the motion-preservation zone towards replicating the asymptomatic spinal segment's ligaments strain profiles and its facet joint force profiles during main motions. This design process aims to minimize the risk for postoperative pain, avoidable degeneration and to restore segmental biomechanics, to prevent adjacent segment effects. Designs with single articulation and with dual articulation (with a mobile insert) were optimized. The bone-implant interface was optimized with the aim to minimize the risk for subsidence and implant migration. The optimizations improved the multi-objective value of the bone-implant interface by 14.6% and that of the motion-preservation zone by 36.1%. Implant migration, the leading cause of revisions, was reduced by 24.8%. With this, we show the potential of simulation-driven implant design optimization for addressing complex clinical challenges.

## 1. Introduction

We move our neck thousands of times per day[1], which can pose a substantial burden on patients when it is connected to pain. Neck and arm pain are a major and growing global burden. Neck pain alone affected upwards of 300 million people for over three months in 2015 globally[2], with close to 35 million years lived with disability, which is a 21% increase compared to 2005[2,3]. Furthermore, neck and arm pain can be accompanied by neurologic deficits. Underlying causes can be degenerative disc disease, disc herniation, spondylosis or facet joint arthrosis leading to myelopathy or radiculopathy. When conservative treatment proves ineffective within a reasonable timeframe, invasive treatment may be considered. Two main surgical treatment options are spinal fusion and spinal arthroplasty.

While in cervical fusion typically a cage with or without an anterior plate immobilize the segment, in arthroplasty movement is preserved using a Total Disc Replacement (TDR) implant. Both fusion and arthroplasty are overall safe and effective[4,5]. Recently, cervical arthroplasty has become increasingly popular, with the number of procedures increasing by 654.2% between 2011 and 2019 (plateauing in 2019-2021, possibly due to the COVID-19 pandemic)[6]. While cervical fusion remains more common, its use has stagnated between 2014 and 2021[6]. One reason for this might be that arthroplasty shows lower revision rates than fusion. A meta-analysis found that in a 7-year follow up the reoperation rate was less than half of that in the fusion group in both index and adjacent levels (TDR: index: ~5%, adjacent: ~4%; fusion: index: ~13%, adjacent: ~11%)[5]. The most common reasons for revision of cervical arthroplasty were migration (23.5%) and neck pain (15.5%), based on the Manufacturer and User Facility Device Experience (MAUDE) database[7] (Figure 1).

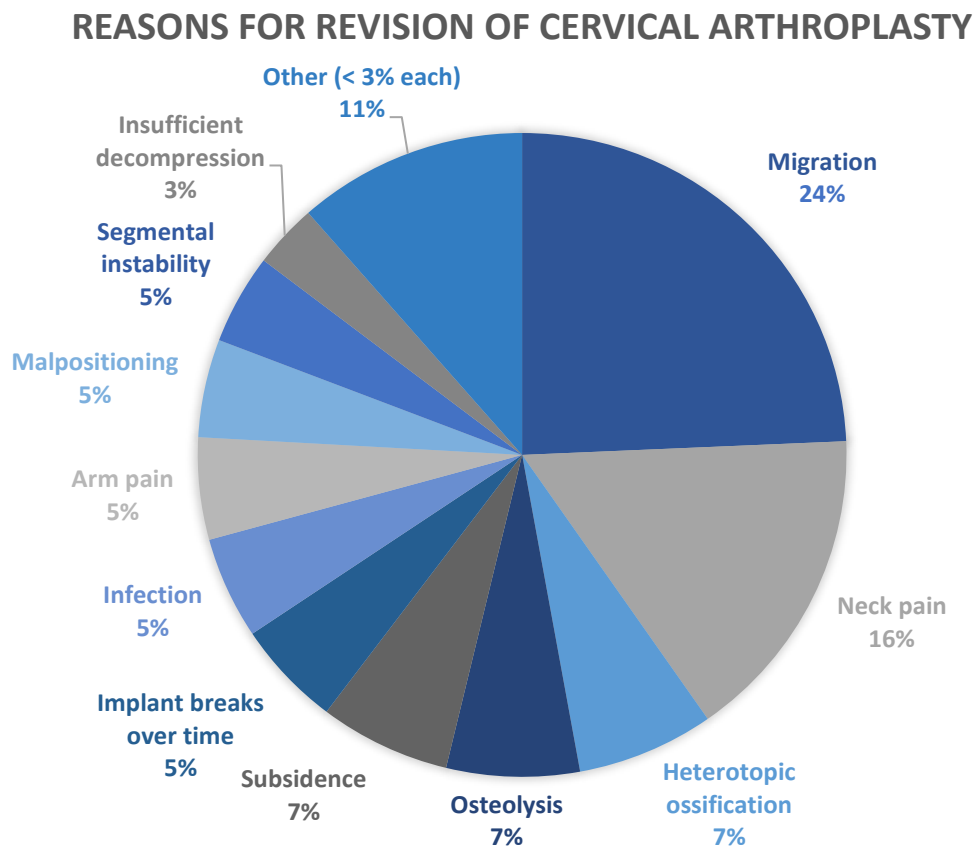

Figure 1: Complications underlying cervical arthroplasty revision cases. Figure based on data provided in Altorfer et al.[7]. Complications that caused less than 3% of revisions each were adjacent segment disease, traumatic dislocation, severe patient injury during surgery, wrong size, device malfunction and allergy.

TDRs have two design zones: a motion-preservation zone and a bone-implant interface. The motion-preservation zone's function is to enable segmental motion, whereas the bone-implant interface's function is to securely connect the TDR to the adjacent vertebrae. Another function of TDRs is to restore the disc height and thereby decompress neural structures to relieve the patient of pain.

TDR design affects motion, however, there is no consensus on which design enables the most physiological motion profiles[8]. It is desirable to enable the remaining spinal structures and central nervous system to perform movements that are as close as possible to what they would be in the asymptomatic condition. Approximating physiological motion has the potential to reduce the risk for adjacent segment effects, postoperative pain and avoidable degeneration.

As the most mobile spinal region, the cervical spine depends on stabilization through ligaments[9]. Facet joints and intervertebral discs (IVDs) also stabilize and guide spinal motion[9]. This interplay of spinal structures and the resulting cervical spine kinematics can be affected by the introduction of a cervical TDR, which ideally functionally replaces the structures that are removed (excised) or cut (dissected) during implantation surgery (IVD, anterior longitudinal ligament (ALL) and typically also the posterior longitudinal ligament (PLL)).

TDR design parameters, placement and orientation likely alter facet joint mechanics at index and adjacent levels[10]. Cadaveric studies found that TDRs might alter facet joint loading[8], and an *in silico study*[11] found that for all four compared TDRs facet joint forces were increased at the index level (2.04 - 3.2 times intact condition) and reduced at adjacent levels (76% - 87% of intact condition). As cartilage degeneration and disc degeneration are associated with altered facet biomechanics[10], this effect is relevant. Facet joint capsule and subchondral bone are highly innervated with nociceptive, mechanoceptive, and proprioceptive nerve endings, meaning that mechanical loading of these structures could cause pain[10]. Being the main stabilizing structures of the facet joints, the capsular ligaments have been suggested as a major source of chronic neck pain[9]. Elongation-induced strains of the capsular ligaments may progress to capsular ligament laxity[9]. Laxity of capsular ligaments might cause cervical radiculopathy, and with that pain, tingling and/or numbness in arms/hands, as hypertrophic facet joint changes might affect nerve roots[9]. The ALL and PLL as well as the ligamentum flavum prevent excessive flexion and extension and aid in stabilizing the cervical spine[9]. As ALL and PLL are commonly excised or dissected during cervical arthroplasty surgery, ideally the TDR could support the remaining ligaments functionally and thereby protect them. The posterior ligaments are innervated with nerve endings and may generate pain when injured[9]. Postoperative pain connected to ligaments and facet joints might not have received the attention it deserves, and this may be partially why TDRs do not perform to their full potential yet.

Neck pain is the second most common reason for revision of cervical arthroplasty, the most common one being migration[7]. An implant design that mechanically resists migration (anteroposterior TDR displacement) is beneficial for all patients initially, but those that have sub-ideal conditions for osseointegration, may rely on it permanently. That migration was the most common reason for revision in seven of the nine FDA-approved TDR models[6] indicates that the bone-implant interface design might not have received sufficient attention and there appears to be potential for improvement.

This study aimed to develop a method to specifically optimize TDRs to reduce the risk for complications associated with current TDRs such as migration, subsidence, and pain related to the facet joints and ligaments. For the bone-implant interface specifically, the objectives were to minimize predicted migration and subsidence. Whereas for the motion-preservation zone, the objectives were to minimize the risk for potential indicators of pain and avoidable degeneration and to replicate intact segmental biomechanics, to reduce the risk for adjacent segment effects. To this end, we created a simulation-driven design optimization framework based on thorough understanding of cervical arthroplasty from an engineering and clinical point of view, advanced finite element simulations, and TDR designs employing an established medical ceramic material. With this, we addressed factors responsible for more than 50% of revisions of cervical arthroplasty.

## 2. Materials and Methods

Computational design optimizations of a TDR's bone-implant interface, and the motion-preservation zones of a single articulating and a dual articulating TDR were performed. Multiple design variables were defined and varied which created various implant geometries that were investigated in numerical experiments (finite element simulations) with respect to design objectives and constraints. Based on the simulation results, the metamodels, neural networks, were trained and used to reduce the design spaces to the most expedient sub-domains. These automated processes were performed iteratively until the termination criteria were reached.

The optimizations were set up in LS-OPT 7.0.2 (LSTC, USA). They used custom Python-scripts that utilized the PythonOCC library[12] for the geometry creation, ANSA v24 (BETA CAE Systems International AG, Switzerland) for meshing, LS-PrePost 4.10 (ANSYS, Inc, USA) for mesh improvement and LS-DYNA R14.1 (ANSYS, Inc, USA) for the finite element simulations. The simulations were run on a large computer cluster (>200,000 CPUs), all other operations on a compute cloud. The best performing designs of the last iterations were selected as the optimized designs.

## 2.1 Spine simulation model

The open-source finite element human body model VIVA+[13] has a detailed cervical spine model including ligaments, intervertebral discs (incl. annulus fibers, annulus ground, nucleus), vertebrae (trabecular bone, cortical bone and endplates), and facet joints (incl. cartilage surfaces, capsular ligament). The cervical spine models segmental moment-rotation curves were previously validated against cadaveric data[14]. The model also contains muscles and can be actuated by them; however, we did not include the muscles in this study.

In the present study, the 50th percentile male standing model was chosen to design an implant that would benefit a substantial proportion of the population. The C5/C6 level was chosen, as it is the level most frequently operated on for both cervical arthroplasty and cervical fusion (51% and 50%)[15]. The C5/C6 functional spinal unit of the VIVA+ model is depicted in Figure 2.

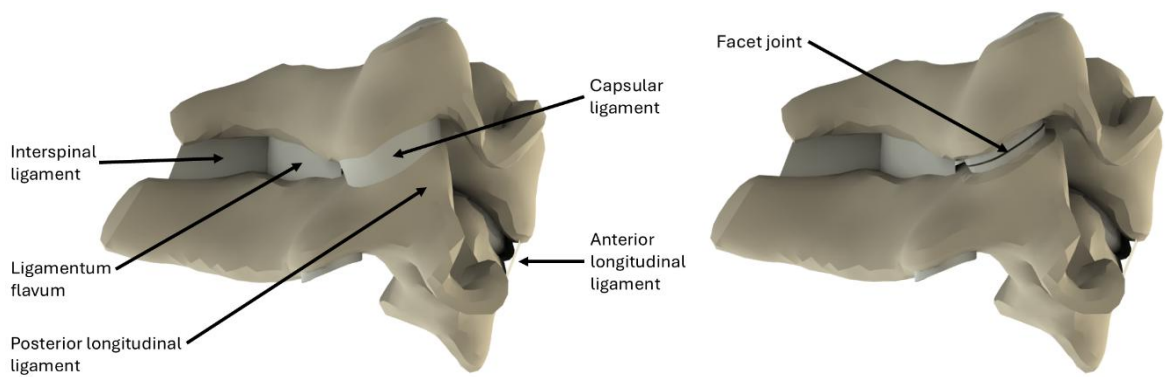

Figure 2: C5/C6 cervical spinal segment of the 50th percentile male VIVA+ FEM model in lateral view with all ligaments (left) and with the capsular ligament being hidden to show the facet joints cartilage surfaces (right). The posterior longitudinal ligament is included but cannot be seen from this perspective.

## 2.2 Baseline implant designs and design concept

The baseline TDR designs consist entirely of zirconia-toughened alumina ceramics and are shown in Figure 3. The rational for the selection of the ceramic material[16], the material model and material parameters of the ceramic material[17], the single articulating bearing design[17] and experimental testing as well as tribological investigations[18] of it are described in previous publications.

There are two baseline designs for the motion-preservation zone: one with a single articulation couple, a ball-in-trough type and a dual articulation version with a biconvex mobile insert, both are shown in Figure 3. The single articulation baseline design is the result of a previous optimization study[17]. The inferior trough of the dual articulation baseline design is identical to the trough of the single articulating design. Since the superior endplate is shorter in anteroposterior direction, a smaller superior trough was used in the baseline design.

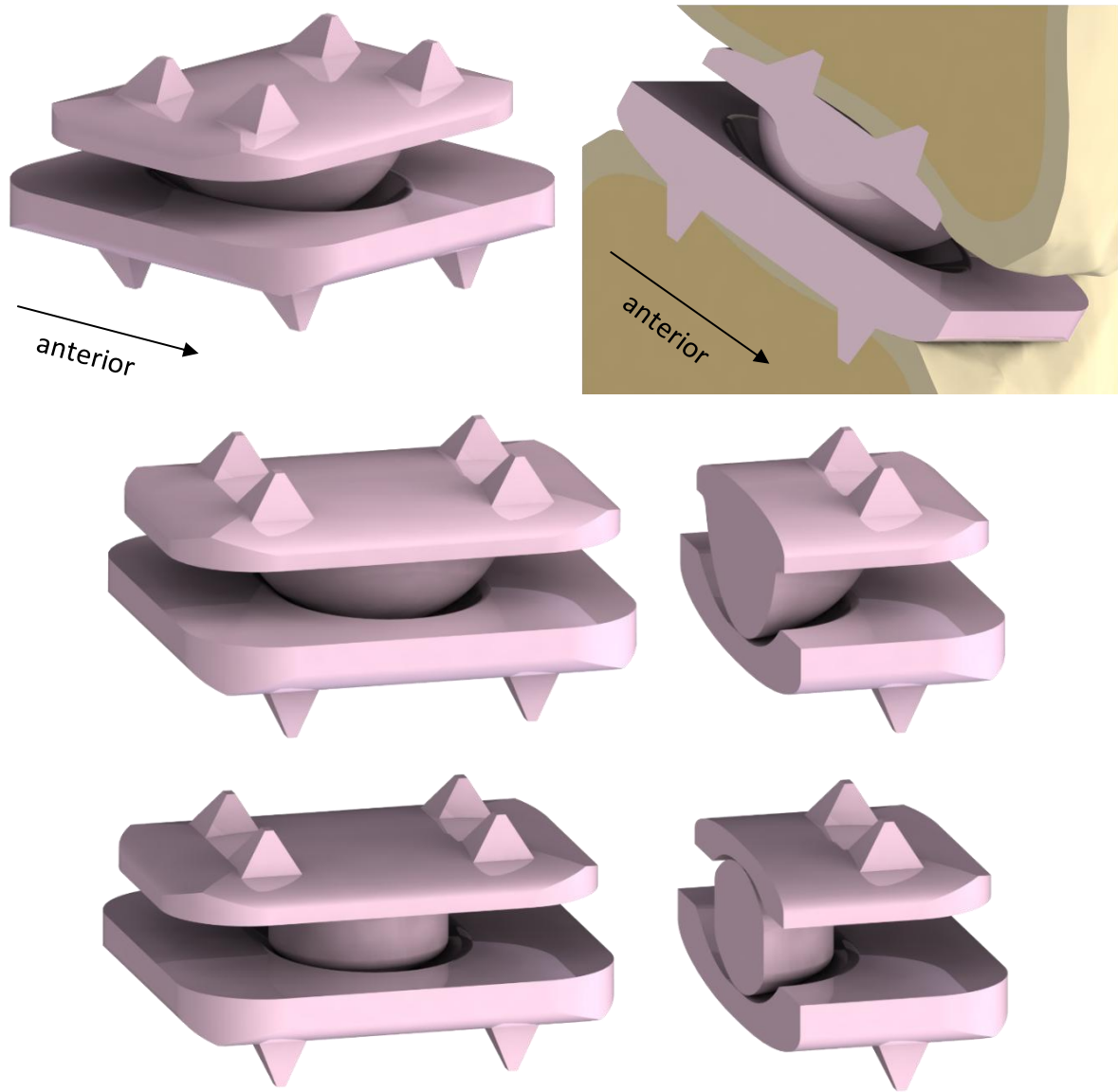

**Figure 3: Top row: Left:** Baseline TDR design of the bone-implant interface and single articulation optimization. **Right:** Baseline TDR design in the bone-models used for the bone-implant interface optimization (detail-view of cross section), the cortical and trabecular bone can be seen.
**Middle row:** Baseline design of the bone-implant interface and single articulation optimization (left) and cross section of it (right).
**Lower row:** Baseline design of the dual articulation optimization (left) and cross section of it (right). This design consists of three parts.

Since keels were linked to heterotopic ossification, likely due to opening of the vertebral cortex[19]; and to vertebral split fractures[19], our baseline design uses spikes which are also easier to implant. Spikes with concave curvatures might create gaps in the bone in front of them following the implantation which may enable micromotions, therefore, we focus on spike tapering. Furthermore, our baseline designs spikes are intended to be comparable to typical current TDR features, and its endplates were designed to have a good match with the vertebral endplates shapes to preserve bone. The corners of the endplates are radiused, and the sides are rounded using asymmetric fillets.

The design rationale for the motion-preservation zone considers the characteristic motion coupling of the lower cervical spine segments, especially coupled flexion/extension-anteroposterior translation. While ball-in-socket designs do not enable translation, designs with moving centers of rotation such as ball-in-trough designs or dual articulating designs with a mobile insert do.

The design space of the single articulation optimization includes ball-in-socket designs and ball-in-trough designs. That of the dual articulation optimization includes designs with two sockets, two troughs or one of each. The centers of the spherical implant parts can be varied in anteroposterior direction in the design optimization process. This shifts the center of the spherical part ("dome"/"ball") and the socket or trough, or, in case of the dual articulating designs the mobile insert and the sockets or troughs. The centers of the spherical parts (centers of the curvatures of the spherical parts of both the single and dual articulation designs) can also be shifted in the craniocaudal direction leading to more or less of the spherical caps being built.

## 2.3 Virtual implantation of TDR into the spine model

For the virtual implantation of the TDR into the spine model, the TDRs were scaled to 75% to match the size of the segment. The IVD, ALL and PLL were removed before the TDR was inserted. As the vertebral endplate is stronger posteriorly than anteriorly, even following burring, posterior implant placement is favored[20].

## 2.4 Bone-implant interface optimization

The optimization of the bone-implant interface aimed to minimize the risks for subsidence and migration. The corresponding optimization problem can be formulated as:

$$\min_{p} \quad w_1 * |d_{subsidence}(p)| + w_2 * |d_{expulsion}(p)| \quad (1a)$$

$$\text{s.t.} \quad p \in \Omega \quad \text{(design space)} \quad (1b)$$

$$\sigma(p) \leq \sigma_{max} \quad \text{(stress constraint)} \quad (1c)$$

$$d_{micro}(p) \leq d_{micro\_max} \quad \text{(micromotion constraint)} \quad (1d)$$

The objective $d_{subsidence}$ refers to the craniocaudal displacement of the C5 vertebra (the superior bone) during simulations of the subsidence load case, and the objective $d_{expulsion}$ refers to anteroposterior displacement of the TDR in the simulations of the expulsion load case. The weights $w_1$ and $w_2$ are the weights of the objectives, the vector p contains the geometric design variables that are subject to the bounds of the design space $\Omega$ (see Figure 8 and Table 1 in the Appendix). These design parameters define the shapes of the endplates, as well as the shapes and positions of the structures for primary fixation.

The multi-objective design optimization used a metamodel-based optimization strategy that was sequential with domain reduction, the workflow is shown in Figure .

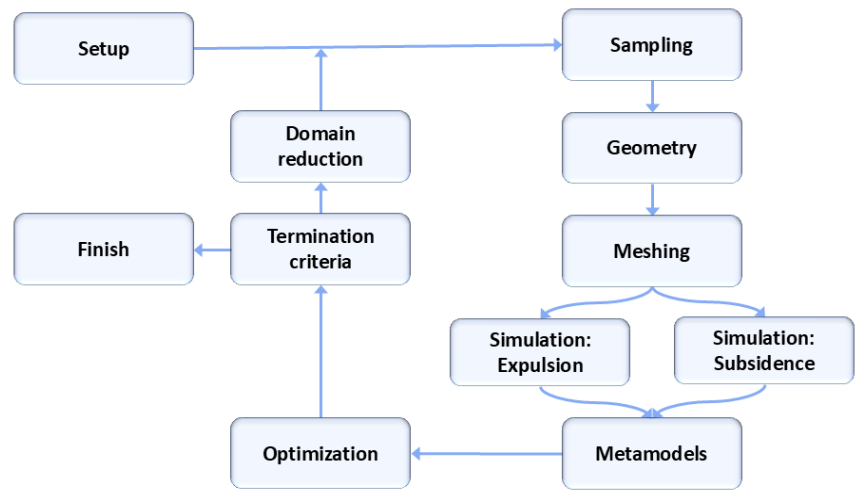

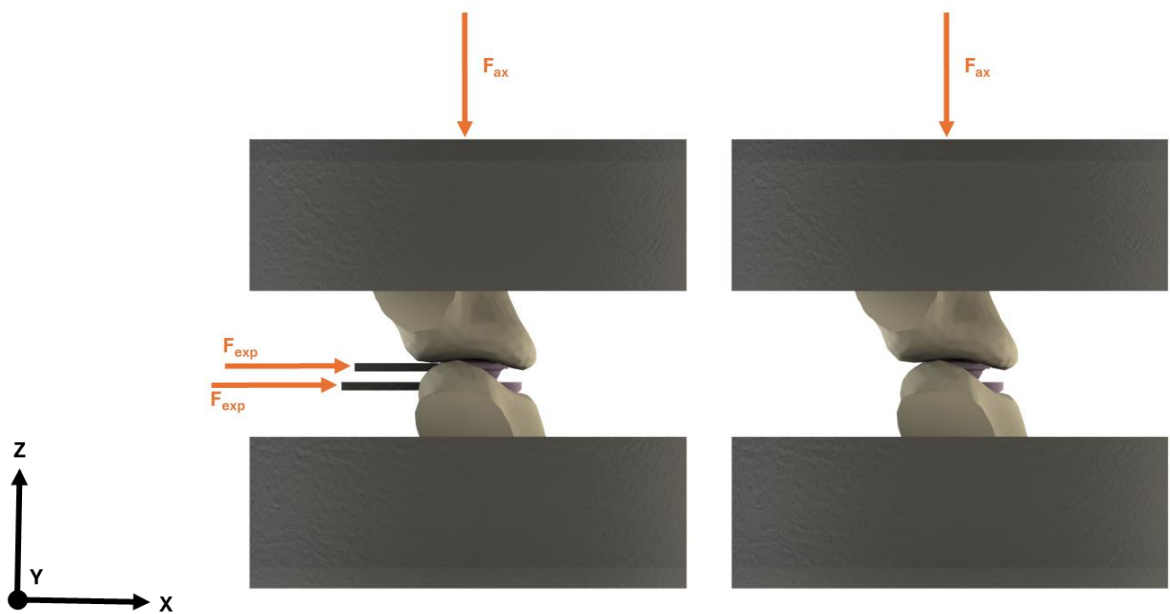

**Figure 4: Upper row: Flowchart for the optimizations of the caudal and cranial (inferior and superior) bone-implant interfaces. This workflow was set up in LS-OPT.**
**Lower row: Expulsion and subsidence load cases. The coordinate system (black) is the same in both cases (x = anterior, z = cranial). The grey cylinders model potting blocks used to apply boundary conditions. The inferior potting is fixed in all six degrees of freedom. The gray blocks model the indenters used in the expulsion-test simulations. The TDR is shown in pink, the vertebral bodies in beige.**

The automated optimization workflow (Figure 4) started with a ***setup stage***, in which the initial design space and the baseline design were defined. In the following ***sampling stage***, design parameter sets were selected. In the ***geometry stage***, the implant geometries were created (Figure 11 in the Appendix) according to the parameter sets and cut out of the vertebral bone where they intersected. The bone and TDR geometries were then meshed in the *meshing stage*. The meshed bones and TDRs were used in the subsequent finite element *simulation stages*, in which the expulsion and subsidence load cases were simulated in parallel. The results of these simulations were used to build surrogate models in the *metamodels stage*. In the *optimization stage*, an optimal design was predicted based on the metamodels, with respect to the objectives and taking the constraints into account. The following *termination stage* checked whether one of the termination criteria was reached, and if they were not reached yet, the optimization entered the **domain reduction stage** in which the design space was reduced to its most expedient part. Based on this reduced design space, the sampling stage was then entered again, and the process was iterated until one of the termination criteria was met, whereupon the optimization *finished*.

The initial design space was high-dimensional (34 design parameters), and complexity of a design optimization grows exponentially with the number of parameters. As the caudal ("lower"/inferior) and cranial ("upper"/superior) bone-implant interfaces design parameters effects are likely (largely) independent of each other, we increased the optimization efficiency by optimizing them separately (two 17 design variable optimizations instead of one with 34). In each of the two optimizations, 125 designs were investigated in each iteration. To give equal importance to both design objectives, we first ran a Design of Experiments (DOE) study with 100 designs to obtain the scale factor that relates to the proportion of the average displacements different designs reached. This scale factor was then used for the weighting of the objectives in the following optimizations of the superior and of the inferior bone-implant interfaces. Following the optimizations, the performance of the full bone-implant interface (superior and inferior) was evaluated.

## Metamodel and Optimization algorithm

Radial basis function neural networks were chosen as the surrogate model, as they can be trained faster than typical (deep) feedforward neural networks[21], and offer a high prediction accuracy[21]. We included points of previous iterations and used space filling point selection, with a *maximin* distance space filling algorithm (maximizing the minimal distance between sampling points (designs)). Hardy's multi-quadratics was selected as the basis function of the artificial neural network.
The optimization used a hybrid optimization algorithm that started with a genetic algorithm that finds an approximate global minimum and refines using a gradient based optimizer that searches for a local minimum; to improve the speed of finding a (global) optimum. This was done with a population size of 100, with 250 generations.
These choices correspond to pre-existing options in LS-OPT.

## Implant geometry and simulations

If fixation-structures geometrically overlapped, they were merged, enabling the generation of serrated keels from overlapping spikes. The thickness of the vertebral endplate is important for the bone-implant interface. Therefore, we modelled the thickness distribution of the cortical shell, including the bony endplate, based on a published study[22], see Figure 12 in the Appendix. We assigned the material models and properties of the cortical and trabecular bone of the respective cortical and trabecular bone models in the VIVA+ model. To reduce simulation time, only the most relevant structures were part of the simulation models while ligaments, muscles and posterior structures were removed. The bone and implant were meshed with tetrahedral elements, the target mesh size of 0.4mm was selected with a mesh sensitivity study. The friction coefficient between bone and implant was conservatively set to $\mu = 0.3$ (commonly used value for bone-implant interfaces[23]) as to rely on mechanical interlocking rather than friction. Although higher friction may be achieved with a ceramic coating ($Al_2O_3$ blasted surface/human trabecular bone $\mu = 0.48$[24]). The coefficient of friction between the ceramic implant parts was chosen as $\mu = 0.16$ based on our previous study[18]. We cut out the bone to match the implant geometry, as it is more conservative than pressing the implant into the bone and thereby creating a strong bony foundation in the bone-implant interface.

The load cases migration (expulsion) and subsidence were defined with the intent to generate more realistic versions of typical *in vitro* tests. Loads were applied to the top surface of the superior potting block and the expulsion blocks. Figure 4 shows the simulation setups used, Figure 13 in the Appendix visualizes the load cases used in the implicit finite element simulations. The main differences compared to typical tests are that our simulations used vertebra models instead of polyurethane foam blocks or metal blocks and loading that focussed on realistic load ranges. Subsidence risk is typically investigated based on the testing standard ASTM F2267[25] (see Figure 15 in the Appendix). Migration is commonly investigated with an expulsion test for which there is no standard, but the Summary of Safety and Effectiveness Data (SSED)[26] report the methodologies and acceptance criteria. A typical expulsion test applies a compressive load to the implant and additionally applies an expulsion-load via an indenter (see Figure 15 in the Appendix). As clinically, one of the TDR parts can migrate more than the other, and to consider possible subluxation (the ball sliding out of the trough) two indenters are used in our simulation setup to capture more potential complications. The simulations ran on 4 CPUs each, with a wall-time of 4hours in case of expulsion, and 2hours in case of subsidence.

## Design constraints

Constraints connected to reliability, manufacturing, implantation, micromotion, anatomy, and design were implemented. Constraint scaling was selected in LS-OPT to normalize the constraints. A material failure constraint $\sigma(p) \leq \sigma_{max}$ was implemented to prevent fracture of the ceramic TDR by limiting maximal principal stresses during the simulations to $\sigma_{max} = 0.3$ GPa. Based on "extra-high strength" zirconia-toughened alumina ceramics biaxial flexural strength given in ISO 6472 Part 2 (2019)[27] and a safety factor of 2. As this study designed for conventional ceramic manufacturing, there were some related constraints. These were implemented as sampling constraints and via the size of the design space.

While low-amplitude micromotions (<40-70μm) can promote bone-remodeling, larger interfacial micromotion (>150 μm) might lead to periprosthetic fibrous tissue instead of osseointegration[24]. We therefore used a micromotion constraint preventing anteroposterior motion of > 150 μm.

Implantability is a crucial aspect for TDRs. Teeth/spikes are typically implanted by compressing the segment using a retainer and thereby creating matching indentations in the bone. As with this approach only a limited amount of force can be used we limited the area of the outer surfaces of the spikes to 2.5mm$^2$ for each bone-implant interface (superior and inferior).

Over-distraction could cause pain, issues with kinematics, and harm of the posterior ligaments and facet joints. Therefore, we limited additional distraction to max. 4mm, meaning that the height of the spikes could be max. 4mm (sum of superior and inferior spikes) so that the segment would temporarily be distracted to at most this height plus the implant height during implantation. As we want the structures for primary fixation to penetrate the vertebral endplate, these structures should be at least 0.5 mm high. The thickness of the inferior vertebral endplate of C5 and the superior endplate of C6 are reported as 0.53-0.75mm, depending on the position on the endplate[22]. As these structures should prevent migration of the TDR without breaking, withstand the loading they experience during implantation, and we aim for conventional ceramics manufacturing, these structures should not be too thin. Therefore, we set a constraint on the minimal length of the spikes' distal edges lengths of 0.25mm. We enforced that each major radius must be larger than its corresponding minor radius. Furthermore, we used a constraint to limit how peripheral the spikes could be so that they did not extend out of the endplates' curvatures.

## Termination criteria

The optimization was set to finish if any of the following termination criteria were reached: a design change tolerance of 1%, an objective function tolerance of 1%, or 50 iterations. The design change tolerance reflects a 1% sintering tolerance in conventional ceramic manufacturing. The termination criteria definitions were already implemented in the software LS-OPT and are given in the Appendix.

## 2.5 Motion-preservation zone optimization

The optimization of the motion-preservation zone aimed to replicate the preoperative maximal principal strains of the ligaments and the facet joint forces throughout motions. To this end, the optimization minimized the difference between the temporal curves of the postoperative condition and the preoperative condition. We performed a multi-objective design optimization with four objectives for each of the four motions we investigated, therefore, the optimization had 16 objectives. Specifically, the maximum principal strains in the capsular ligament, interspinal ligament and ligamentum flavum, and the resultant force between the facet joint cartilage in the load cases: flexion, extension, lateral bending and axial rotation.

We formulated the corresponding optimization problem as:

$$\min_{p} \sum_{i=1}^{16} w_i * MSE_i(p) \qquad (2a)$$

$$\text{s.t.} \quad p \in \Omega \qquad \text{(design space)} \qquad (2b)$$

$$\sigma(p) \leq \sigma_{max} \qquad \text{(stress constraint)} \qquad (2c)$$

Where $MSE_{1-16}$ are the normalized mean square errors between the temporal curves of the preoperative and postoperative ligament maximal principal strains and facet joint contact forces in all load cases. All objectives are weighed equally ($w_i = 1$). The vector p contains the geometrical optimization variables (see Figure 9, Figure 10, Table 2 and Table 3 in the Appendix). The design space $\Omega$ imposes direct bounds on these variables transcribing constraints.

Figure 5 shows the preoperative model which was used to record the target curves for the design objectives. These curves were output through LS-OPT at 10ms intervals. Load application is shown in Figure 5, specifically oriented to a local coordinate system in the IVD.

The workflow of the motion-preservation zone optimization, shown in Figure , is largely similar to that of the bone-implant interface optimization that was described previously.

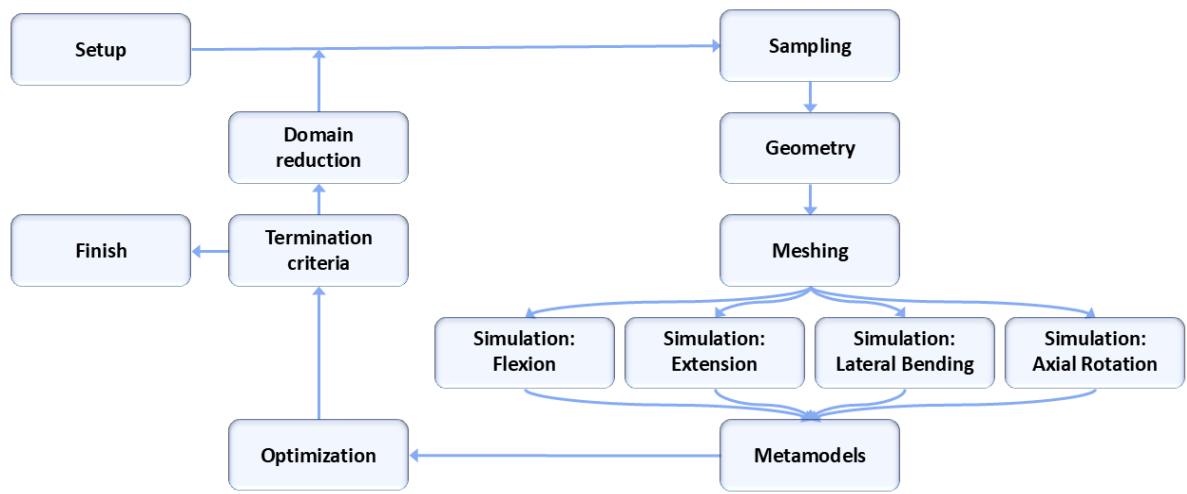

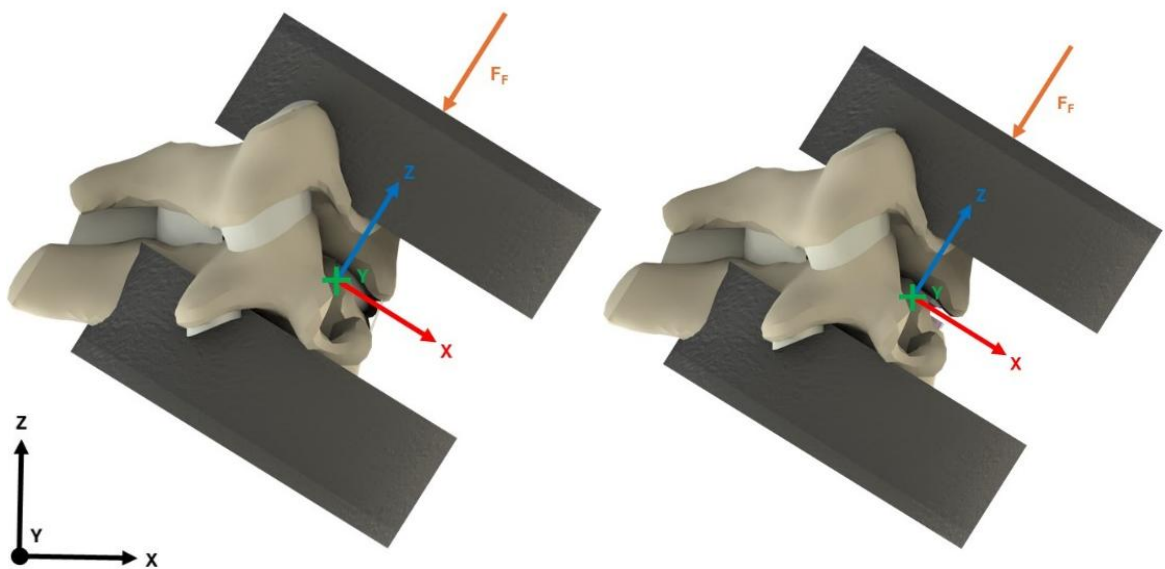

Figure 5: Upper row: Flowchart of the motion-preservation zone optimizations. This workflow was set up in LS-OPT.
Lower row: Left: Preoperative model used to record target curves for the design objectives. The grey cylinders model potting blocks used to apply boundary conditions. The local coordinate system for moment application (blue, green and red) and the load application point of the follower load (orange) used in the simulations of the articulation optimization are shown as well. The global coordinate system is shown in black. The inferior potting is fixed in all six degrees of freedom. Right: Postoperative model used in the simulations of the motion-preservation zone optimization. The IVD, ALL and PLL were removed, and a TDR was added, compared to the preoperative model. The segment is modeled at the angle it has in the model/patient relative to the global coordinate system (in the simulations for the bone-implant interface optimization, this was adjusted to be more inline with *in vitro* tests, here it is kept realistically).

The optimization of the single articulation design had 7 design parameters, and the optimization of the TDR with a mobile insert, and therefore dual articulation, had 14 design parameters, both with additional dependent design parameters. The single articulation design optimization investigated 30 designs per iteration; the dual articulation design optimization investigated 100 designs per iteration as a design optimizations complexity grows exponentially with the number of design variables. The single and dual articulation optimizations design parameters and design spaces are given in Figure 9, Figure 10, Table 5 and Table 6 in the Appendix. While some of the basic motions (flexion, extension, lateral bending, axial rotation) may be performed more often than others in daily living activities, they were weighed equally in the optimization, as issues during any of them would be problematic but also would likely lead to issues through related coupled motions. The single and dual articulation optimizations were run in parallel.

## Metamodel and optimization algorithm

Similar to the bone-implant interface optimizations, radial basis function neural networks were used as metamodels for the same reasons as described before. The optimization algorithm was identical to that used in the bone-implant interface optimization.

## Implant geometry and simulations

The radial clearance (difference between the spherical parts radius and the sockets or troughs mediolateral radius) was 0.07 mm. In case of dual articulation designs, each articulation interface was given a radial clearance of 0.07 mm. To avoid the risk of edges chipping and to facilitate manufacturing, fillets were used in the dual articulation version to round off the mobile insert where it transitions between spherical caps and cylinder. In the single articulation version, there was a fillet between the spherical part and the endplate, and in both design types, there were fillets on the edges of the troughs, as can be seen in Figures 9 and 10 in the Appendix.

In the simulation setup for the motion preservation zone, the segmental model of the original VIVA+ model was used without any changes. The posterior structures were preserved, as they have significant effects on segmental motion. As this optimization was not concerned with the bone-implant interface, it was not necessary to adjust the vertebra model or to cut out the implant geometry from the bone and re-mesh the spinal model. Instead, the bones and the implant-parts were connected through a constraint, and the TDRs were modeled without structures for primary fixation. The target mesh size of the TDR was selected as 0.4mm based on mesh sensitivity studies for the single and dual articulating designs. The simulation setup used to run the explicit finite element simulations is shown in Figure 5.

The main motions of the spine were simulated: flexion, extension, lateral bending and axial rotation. Inspired by a previous study[28], the load cases were chosen as a compressive preload of 73.6N combined with 1.8Nm in flexion, lateral bending or axial rotation. In extension, the same compressive preload was applied, however only 1Nm was applied to prevent lift-off. The compressive load was applied as a follower load perpendicular to the top surface of the upper potting, whereas the moment was applied around a coordinate system in the center of the IVD/TDR. Figure 14 in the Appendix visualizes the corresponding load cases and describes them in more detail. The simulations ran on 12 CPUs each with a wall-time of 13hours. To determine the positioning of the compressive load relative to the spinal segment, prior to the optimization, a simulation was run in which only this load was applied to the preoperative model and moved in the anteroposterior direction until only minimal rotation resulted (<0.25deg).

Motion-coupling of multiple rotations or rotations and translations occurs in the lower cervical spine *in vivo*. In this study, we chose to use standard load cases and let the model produce the motion coupling. To investigate whether the motion-coupling of the preoperative model was realistic, we output the movements of the superior and inferior endplates of C5 in the local coordinate system that is also used for the moment application, see Figure 5, for the most relevant motions. A comparison of these values to a cadaveric study and *in vivo* data is given in Appendix 6.

## Design constraints

Design constraints related to reliability, manufacturability, and function were implemented. Some of the following constraints were implemented as typical constraints, so that for example following the simulations a stress constraint was evaluated, while others were implemented as sampling constraints, so that for example design parameter sets that would create a mobile insert that is narrower than it is tall would not be selected in the first place. In the dual articulating version, the mobile insert had to be wider than tall so that it would not flip on its side, reducing implant height and compromising function. An impingement constraint was used to prevent component interference in flexion and extension. The material failure constraint $\sigma(p) \leq \sigma_{max}$ for the ceramic implant material was implemented as in the bone-implant interface optimization ($\sigma_{max}$ = 0.3 GPa). Constraint scaling was selected in LS-OPT to normalize the constraints.

## Termination criteria

Reaching any of the following termination criteria was sufficient to end the optimization: a design change tolerance of 1%, an objective function tolerance of 1% or 50 iterations.

# 3. Results and Discussion

The aim of this study was to develop a method to computationally optimize the design of a TDR to minimize the risk of complications of current TDRs. More specifically, we aimed to optimize the bone-implant interface to minimize the risks for migration and subsidence; and we aimed to optimize the motion preservation zone to replicate the asymptomatic segmental behavior with the intention to minimize the risk of pain and avoidable degeneration, related to loading of anatomical structures, and to prevent adjacent segment effects. To this end, we conducted finite element simulation-driven design optimizations of a cervical TDRs bone-implant interface and motion preservation features. s

That the termination criteria were met after relatively few iterations is likely connected to the high numbers of designs we investigated per iteration allowing the neural networks to train on large datasets already within the first iterations.

## 3.1 Bone-implant interface

After exploring 250 designs (two iterations) in case of the superior and 750 designs (six iterations) in case of the inferior bone-implant interface optimization, the objective function termination criteria were reached. The optimized design is shown in Figure 6, its design variables values are given in Table 1 in the Appendix. The multi-objective value improved by 14.6% compared to the baseline design. Specifically, 3.9% for the subsidence objective and 24.8% for the expulsion objective.

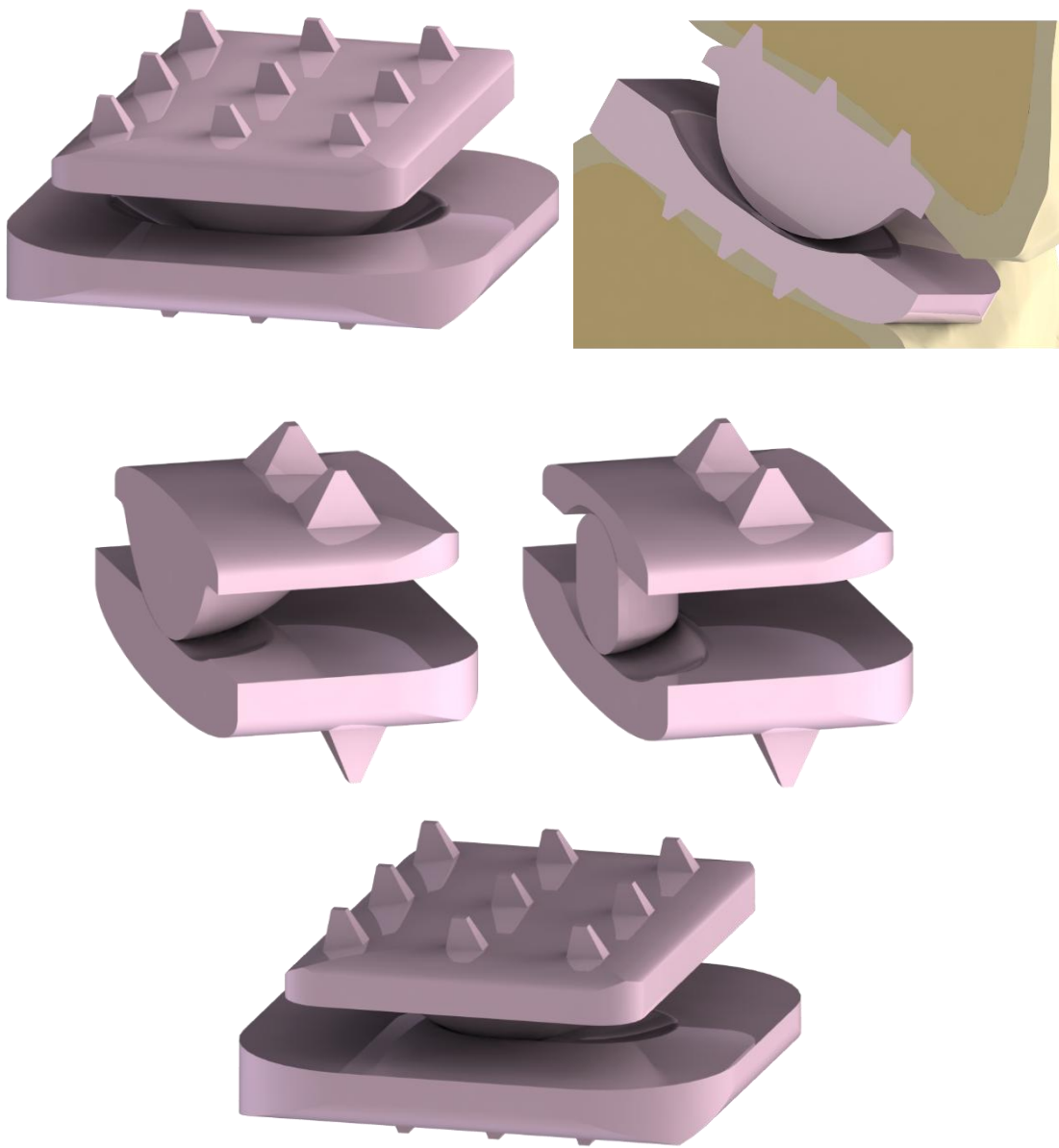

**Figure 6: Top row: Optimized bone-implant interface design. Left: Full TDR, right: uneven cross section revealing the interactions of the cortical and trabecular bone with the TDR.
Middle row: Optimized design of the single (left) and dual (right) articulation motion-preservation**

**zone with the baseline bone-implant interface design (cross sections).
Lower row: TDR with optimized bone-implant interface and optimized single articulating motion preservation zone design.**

The optimization process reduced the objectives (displacements) considerably, suggesting that the process aided in reducing the risk for implant migration and subsidence. The 24.8% improvement in terms of migration is especially relevant, as migration was the most common cause of revision in seven of the nine investigated TDRs in an analysis of the MAUDE database[7]. The expulsion testing acceptance criterion used in a Summary of Safety and Effectiveness Data (SSED) file was ≤3mm movement under 20N force[29]. We used double the force and the optimized design displaced by <0.01mm, indicating that it would likely have passed this test. The optimization of the superior bone-implant interface reached one of the termination criteria within fewer iterations than the optimization of the inferior bone-implant interface, underscoring the efficiency of optimizing them separately. This may reflect a higher complexity of the inferior bone-implant interface optimization problem. The superior endplate, with its shorter length in anteroposterior direction, fits well into the superior vertebral endplate's curvature and more of the superior vertebral endplate remains.

The anteroposterior lengths of the fixations were reduced through the optimization, likely as the length did not contribute additional resistance to expulsion but did lead to larger recesses in the vertebral endplate. The optimized bone-implant interface has a higher amount of primary fixation structures than the baseline design. Multiple smaller openings in the vertebral endplate may reduce its compressive strength less than fewer but larger openings. The vertebral endplates had increased surface area but reduced volume following the virtual implantation of the optimized design compared to the baseline design. This indicates that the remaining endplate shell surface area may be of greater relevance than the volume.

The corners of the optimized design are not very rounded. This is in line with literature, which has claimed that to reduce subsidence risk, a cervical TDR needs to be rather rectangular than round to utilize the lateral trabeculae that are radially oriented and as much as possible of the vertebral endplate[30]. A study on the mineralization distribution of cervical endplates found the highest mineralization in peripheral zones and lower mineralization centrally[31].

Multiple factors may play a role in the bone-implant interface design, such as the thickness and compressive strength distributions across the vertebral endplate. While anchoring structures for primary fixation in areas of thick and strong vertebral endplates, the loss of the endplate where the structures for primary fixation are located or where endplate preparation would remove it to fit the implant, could be detrimental in terms of subsidence. In Figure 6, it is visible that the optimized designs inferior fixation structures anchored into the central part of the vertebral endplate, preserving much of the remaining continuity of the vertebral endplate.

### 3.2    Motion-preservation zone

The optimization of the single articulating design reached the objective function criterion after exploring 120 designs (four iterations), the optimization of the dual articulating design reached it after exploring 900 designs (nine iterations). The optimized designs are shown in Figure 6, their design variables values are given in Table 2 and Table 3 in the Appendix. Their multi-objective values improved by 36.1% for the single articulating design compared to its baseline design and by 16.9% for the dual articulating design compared to its baseline design. The multi-objective value of the optimized single articulating design was 30.1% lower than that of the optimized dual articulating design.

The considerable reduction of the multi-objective values (curve matching) suggests that the optimization aided in generating implant designs that are able to more closely replicate asymptomatic biomechanics. This high-dimensional non-linear problem would have been challenging to optimize without such an automated pipeline.

The highest peak maximal principal strains in ligaments were found for the interspinal ligament during flexion. For the optimized designs they were 65.7% for the single articulating design and 63.04% for the dual articulating design. These values are in proximity to the values of the preoperative condition, where it was 63.8%. The influence of the anteroposterior location of the center of the spherical part on the interspinal ligaments maximal principal strain is shown in Figure  alongside the preoperative (target) curve and a fringe plot of this measure in the preoperative simulation and with the optimized single articulating design. This design parameter is especially relevant, as in some single articulating TDRs, the center of the spherical part is positioned more posteriorly which is reasoned with the flexion/extension center of rotation of the segment being located more posteriorly. However, in the present study, the optimized design does not have a strongly posteriorly placed sphere center (the rather posterior position of the implant adds to the positioning within the TDR, however). The plot of the effect of this parameter on the interspinal ligaments maximal principal strain throughout flexion illustrates the relevance of this design parameter. A previous study[32] reported that anteroposterior positioning of the center of rotation of a ball-and-socket TDR influenced flexion and extension rotations. It is unsurprising, that a design parameter that can affect the amount of rotation might also affect interspinal ligament strain during flexion. In the present study, we ran global sensitivity analyses (Sobol) following the optimizations and found that the design parameter with the highest effect overall (all 16 design objectives combined) for the single articulation design was the anteroposterior location of the center of the spherical part, and for the dual articulating design it was the anteroposterior position of the mobile insert. The anteroposterior positioning of the spherical part and initial positioning of the mobile insert affect the positioning of the troughs/sockets. During motion, the positioning of the insert can change without any changes to the parts' geometries, of course. This is an example of how the results of design optimization can be used to gain a better understanding of complex functional relationships within such systems. Our high-dimensional nonlinear problems can be broken down into simpler insights into the effects of variables as well as how the interaction of variables affect outcomes.

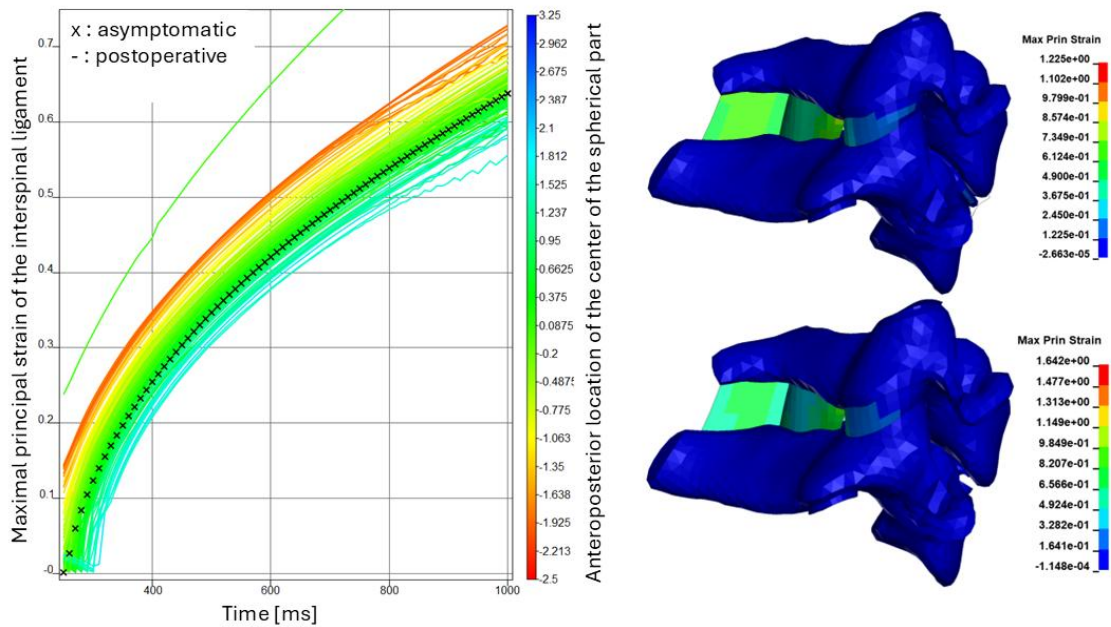

Figure 7: Left: Plot of the interspinal ligaments maximal principal strain over time throughout flexion in the single articulating design optimization (simulations). Individual curves are colour coded according to the anteroposterior location of the center of the spherical part. The target curve is shown with the crosses (asymptomatic condition, maximal principal strain of the interspinal ligament during flexion).
Right: Flexion simulations with the preoperative model and that with the optimized single articulating design visualize the interspinal ligaments maximal principal strains at the end of the simulations.

Previous *ex vivo* and *in silico studies* found that TDRs might alter facet joint loading[8,11]. One study found[11] that all four *in silico* investigated TDRs lead to increased facet joint forces of 2.04 - 3.20 times the intact condition in extension. In the present study, the maximal facet joint forces in extension were 0.81-1.3 times the intact condition in extension for the baseline and optimized single and dual articulating designs. Their preoperative model performed 6.0° of extension, ours performed 6.16°, which is comparable despite differences in modelling and loading. Their hybrid loading approach might be connected to their findings differing from ours, as they report increased ranges of motion in the index level by 5.6% - 31.6%[11]. A comparison of the facet joint forces of our preoperative simulations to a cadaveric study can be found in Appendix 7.

That the optimization of the single articulating design led to a short trough posteriorly and low torus radius posteriorly might be connected to the removal of the ALL (as part of the virtual implantation) which might otherwise cause hypermobility (extension can be achieved too easily) and with that increased loading of the facet joints. Our optimization captures effects like this and adjusts the implant design to prevent or minimize such complications.

. Both optimized inferior troughs are rather shallow; to increase safety, it would be possible to add design features that prevent subluxation.

The design that combines the best performing optimized articulation and the optimized bone-implant interface is shown in Figure 6.

## 3.3 Limitations

General limitations of finite-element simulations apply. As generally in optimization over a (partially) continuous design-space, it is possible that a local, rather than a global optimum was found, and an even better performing design could be possible.

Subsidence can be described as a combination of translational and rotational subsidence (implant sinks in more anteriorly than posteriorly). Rotational subsidence is likely affected by the facet joints. While we modeled the inhomogeneous thickness distribution of the vertebral endplate and endplate preparation, and thereby the inhomogeneous compressive strength distribution, not modelling the facet joints when investigating subsidence is a limitation.

The rather short anteroposterior length of the superior endplate might enable a larger range of motion without impingement and aid in anchoring the TDR in the superior vertebral endplates curvature and it might be seen as an advantage in terms of bone removal for endplate preparation. It might, however, also be seen as a limitation in terms of reduced surface area for subsidence resistance and osseointegration and heterotopic ossification prevention. This length was not a design variable, but a constant value instead.

There may be aspects that limit generalizability across patients: ligament properties may vary with degeneration and between males and females, and facet joint angles might differ between patients or even between the left and right side of a patient's spinal segment (facet tropism). Furthermore, bone quality and vertebral body aspect ratio may vary between patients and there could be level-specific differences in biomechanics. However, the aim of the present study was the method, not the implant design.

## 4. Conclusion

This study demonstrated that computational design optimization can be used to modify TDR geometries towards targets relevant for reducing complications of current TDRs. Optimization objectives were defined to allow the remaining spinal structures to replicate asymptomatic biomechanics and to resist migration and subsidence. The results provide insights in an area where currently there is no consensus on the optimal design. To the best of our knowledge, this is the first study to optimize a TDR's bone-implant interface, and multiple bearing design types for articulation. In addition, this is the first study to optimize a TDR design over such a high-dimensional design space.

The combination of advanced biomechanical simulation models, clinical data on complications of current devices, an advanced ceramic material and design optimization may enable development engineers to incorporate physiological motions and standard lab tests into a partially automated development process and thereby save time and money while speeding up development and obtain optimized implants. The present study illuminates the potential of design driven by clinical need and powered by artificial intelligence.

## Acknowledgement

Funding: European Union's Horizon 2020 programme: Marie Skłodowska-Curie grant agreement No 812765. The authors would like to sincerely thank Jobin John from Chalmers University of Technology, Sweden for his support, the discussions and for sharing simulations that we used as orientation for our simulations.

## References

1. Cobian, D. G., Sterling, A. C., Anderson, P. A. & Heiderscheit, B. C. Task-specific frequencies of neck motion measured in healthy young adults over a five-day period. *Spine (Phila. Pa. 1976).* **34**, E202-7 (2009).
2. Hurwitz, E. L., Randhawa, K., Yu, H., Côté, P. & Haldeman, S. The Global Spine Care Initiative: a summary of the global burden of low back and neck pain studies. *Eur. Spine J.* **27**, 796–801 (2018).
3. Vos, T. *et al.* Global, regional, and national incidence, prevalence, and years lived with disability for 310 diseases and injuries, 1990–2015: a systematic analysis for the Global Burden of Disease Study 2015. *Lancet* **388**, 1545–1602 (2016).
4. Robertson, D. M. *et al.* Cervical Disc Arthroplasty: Rationale, Designs, and Results of Randomized Controlled Trials. *Int. J. Spine Surg.* **18**, 258–276 (2024).
5. Badhiwala, J. H., Platt, A., Witiw, C. D. & Traynelis, V. C. Cervical disc arthroplasty versus anterior cervical discectomy and fusion: a meta-analysis of rates of adjacent-level surgery to 7-year follow-up. *J. Spine Surg.* **6**, 217–232 (2020).
6. Singh, M. *et al.* Anterior cervical discectomy and fusion versus cervical disc arthroplasty: an epidemiological review of 433,660 surgical patients from 2011 to 2021. *Spine J.* **24**, 1342–1351 (2024).
7. Altorfer, F. C. S. *et al.* Reasons for Revision Surgery After Cervical Disk Arthroplasty Based on Medical Device Reports Maintained by the US Food and Drug Administration. *Spine (Phila. Pa. 1976).* **49**, 1417–1425 (2024).
8. Pisano, A. & Helgeson, M. Cervical disc replacement surgery: biomechanical properties, postoperative motion, and postoperative activity levels. *Curr. Rev. Musculoskelet. Med.* **10**, 177–181 (2017).
9. Steilen, D., Hauser, R., Woldin, B. & Sawyer, S. Chronic Neck Pain: Making the Connection Between Capsular Ligament Laxity and Cervical Instability. *Open Orthop. J.* **8**, 326–345 (2014).
10. Jaumard, N. V., Welch, W. C. & Winkelstein, B. A. Spinal Facet Joint Biomechanics and Mechanotransduction in Normal, Injury and Degenerative Conditions. *J. Biomech. Eng.* **133**, 071010 (2011).
11. Purushothaman, Y. *et al.* A Comparison Study of Four Cervical Disk Arthroplasty Devices Using Finite Element Models. *Asian Spine J.* **15**, 283–293 (2021).
12. Paviot, T. pythonocc. (2022) doi:10.5281/zenodo.3605364.
13. John, J., Klug, C., Kranjec, M., Svenning, E. & Iraeus, J. Hello, world! VIVA+: A human body model lineup to evaluate sex-differences in crash protection. *Front. Bioeng. Biotechnol.* **10**, (2022).
14. Jobin D. John, I. P. A. Putra, J. I. Finite Element Human Body Models to study Sex-differences in Whiplash Injury: Validation of VIVA+ passive response in rear-impact. in *IRCOBI conference* 215–226 (2022).
15. Wahbeh, J. M. *et al.* Combining All Available Clinical Outcomes on Cervical Disc Arthroplasty: A Systematic Review and Meta-Analysis. *J. Orthop. Orthop. Surg.* **3**, 1–16

(2022).

16. Kölle, L., Ignasiak, D., Ferguson, S. J. & Helgason, B. Ceramics in total disc replacements: A scoping review. *Clin. Biomech.* **100**, 105796 (2022).
17. Kölle, L. *et al.* Optimization of a bearing geometry for a cervical total disc replacement. *Front. Bioeng. Biotechnol.* **13**, (2025).
18. Kölle, L. *et al.* Advanced preclinical testing of a design-optimized ceramic bearing for a cervical total disc replacement. *PLOS ONE, in press* (2026).
19. Mehren, C. *et al.* Implant Design and the Anchoring Mechanism Influence the Incidence of Heterotopic Ossification in Cervical Total Disc Replacement at 2-year Follow-up. *Spine (Phila. Pa. 1976).* **44**, 1471–1480 (2019).
20. Cheng, C.-C. *et al.* Loss of Cervical Endplate Integrity Following Minimal Surface Preparation. *Spine (Phila. Pa. 1976).* **32**, 1852–1855 (2007).
21. STANDER, N. *et al. LS-OPT® User's Manual A DESIGN OPTIMIZATION AND PROBABILISTIC ANALYSIS TOOL FOR THE ENGINEERING ANALYST*. (2020).
22. Panjabi, M. M., Chen, N. C., Shin, E. K. & Wang, J.-L. The Cortical Shell Architecture of Human Cervical Vertebral Bodies. *Spine (Phila. Pa. 1976).* **26**, 2478–2484 (2001).
23. Ozturk, H., Nair, P. B. & Browne, M. Computational assessment of the coefficient of friction on cementless hip replacement stability. *Comput. Methods Biomech. Biomed. Engin.* **14**, 209–210 (2011).
24. Gao, X., Fraulob, M. & Haïat, G. Biomechanical behaviours of the bone–implant interface: a review. *J. R. Soc. Interface* **16**, 20190259 (2019).
25. ASTM. *ASTM F2267: Standard Test Method for Measuring Load Induced Subsidence of Intervertebral Body Fusion Device Under Static Axial Compression*. (2011).
26. FDA. Premarket approval. *https://www.accessdata.fda.gov/scripts/cdrh/cfdocs/cfPMA/pma.cfm* vol. 502 https://www.accessdata.fda.gov/scripts/cdrh/cfdocs/cfPMA/pma.cfm (2012).
27. ISO, I. O. for S. ISO 6474-2:2019 Implants for surgery — Ceramic materialsPart 2: Composite materials based on a high-purity alumina matrix with zirconia reinforcement. (2019).
28. Goel, V. K. & Clausen, J. D. Prediction of Load Sharing Among Spinal Components of a C5-C6 Motion Segment Using the Finite Element Approach. *Spine (Phila. Pa. 1976).* **23**, 684–691 (1998).
29. Summary of safety and effectiveness data (SSED) Simplify® Cervical Artificial Disc. https://www.accessdata.fda.gov/cdrh_docs/pdf20/P200022B.pdf (2020).
30. Link, H. D., McAfee, P. C. & Pimenta, L. Choosing a cervical disc replacement. *Spine J.* **4**, S294–S302 (2004).
31. Müller-Gerbl, M., Weißer, S. & Linsenmeier, U. The distribution of mineral density in the cervical vertebral endplates. *Eur. Spine J.* **17**, 432–438 (2008).
32. Galbusera, F., Anasetti, F., Bellini, C. M., Costa, F. & Fornari, M. The influence of the axial, antero-posterior and lateral positions of the center of rotation of a ball-and-socket disc prosthesis on the cervical spine biomechanics. *Clin. Biomech.* **25**, 397–401 (2010).
33. Moroney, S. P., Schultz, A. B., Miller, J. A. A. & Andersson, G. B. J. Load-displacement properties of lower cervical spine motion segments. *J. Biomech.* **21**, 769–779 (1988).
34. Lindenmann, S., Tsagkaris, C., Farshad, M. & Widmer, J. Kinematics of the Cervical Spine Under Healthy and Degenerative Conditions: A Systematic Review. *Ann. Biomed. Eng.* **50**, 1705–1733 (2022).

35. Cobian, D. G., Daehn, N. S., Anderson, P. A. & Heiderscheit, B. C. Active Cervical and Lumbar Range of Motion During Performance of Activities of Daily Living in Healthy Young Adults. *Spine (Phila. Pa. 1976).* **38**, 1754–1763 (2013).
36. Zhao, X. & Yuan, W. Biomechanical analysis of cervical range of motion and facet contact force after a novel artificial cervical disc replacement. *Am. J. Transl. Res.* **11**, 3109–3115 (2019).

# Appendix

## Appendix 1: Design variables, their ranges and values

Bone-implant interface

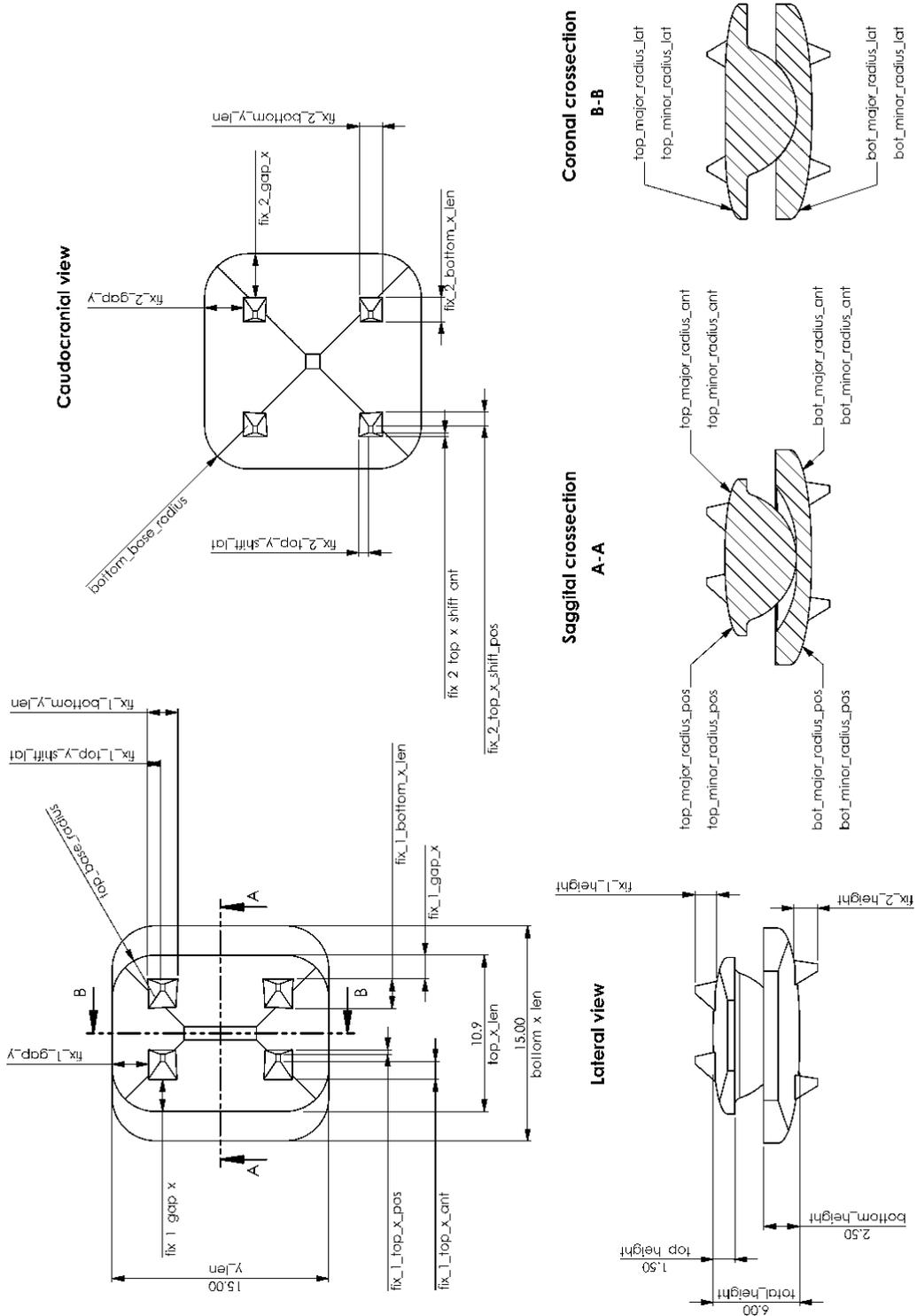

**Figure 8:** The design variables of the bone-implant interface shown on the baseline design. Constant lengths and heights of the bone-implant interface are indicated as well.

**Table 1:** Design space of the bone-implant interface optimization. The parameters regarding the number of fixations are discrete parameters, the others are continuous. The x-direction corresponds to the anteroposterior direction and the y-direction refers to the mediolateral direction in this case. The major radii refer to the horizontal and the minor radii refer to the vertical radii of the asymmetrical fillets; this naming was chosen similar to ellipses minor and major radii.

| Design variable | Min Value | Max Value | Set, if discrete | Baseline design | Optimized design |
|---|---|---|---|---|---|
| *Inferior endplate: bottom_* | | | | | |
| base_radius | 0.25 | 5.0 | | 3.0 | 4.09 |
| major_radius_anterior | 0.25 | 7.0 | | 7.0 | 3.78 |
| major_radius_lateral | 0.25 | 7.0 | | 7.0 | 6.28 |
| major_radius_posterior | 0.25 | 7.0 | | 7.0 | 2.67 |
| minor_radius_anterior | 0.25 | 2.0 | | 1.5 | 1.92 |
| minor_radius_lateral | 0.25 | 2.0 | | 1.5 | 1.66 |
| minor_radius_posterior | 0.25 | 2.0 | | 1.5 | 0.34 |
| *Superior endplate: top_* | | | | | |
| base_radius | 0.25 | 3.0 | | 3.0 | 1.30 |
| major_radius_anterior | 0.25 | 5.0 | | 5.0 | 1.97 |
| major_radius_lateral | 0.25 | 5.0 | | 5.0 | 1.80 |
| major_radius_posterior | 0.25 | 5.0 | | 5.0 | 3.53 |
| minor_radius_anterior | 0.25 | 1.0 | | 1.0 | 0.41 |
| minor_radius_lateral | 0.25 | 1.0 | | 1.0 | 0.51 |
| minor_radius_posterior | 0.25 | 1.0 | | 1.0 | 1.00 |
| *Fixations on the superior endplate: fix_1_* | | | | | |
| number_x | 2 | 3 | ∈ {2, 3} | 2 | 3 |
| number_y | 2 | 3 | ∈ {2, 3} | 2 | 3 |
| Height | 0.5 | 2.0 | | 1.25 | 0.89 |
| bottom_x_len | 1.0 | 3.0 | | 3.0 | 1.89 |
| bottom_y_len | 1.0 | 3.0 | | 3.0 | 1.13 |
| gap_x | 0.0 | 3.0 | | 1.5 | 0.83 |
| gap_y | 0.0 | 5.0 | | 2.0 | 1.81 |
| top_x_shift_ant | 0.0 | 2.5 | | 0.5 | 1.17 |
| top_y_shift_lat | 0.0 | 1.5 | | 1.35 | 0.39 |
| top_x_shift_pos | 0.0 | 2.5 | | 2.0 | 0.29 |
| *Fixations on the inferior endplate: fix_2_* | | | | | |
| number_x | 2 | 3 | ∈ {2, 3} | 2 | 3 |
| number_y | 2 | 3 | ∈ {2, 3} | 2 | 2 |
| Height | 0.5 | 2.0 | | 1.25 | 0.77 |
| bottom_x_len | 1.0 | 3.0 | | 3.0 | 2.65 |
| bottom_y_len | 1.0 | 3.0 | | 3.0 | 2.58 |
| gap_x | 0.0 | 5.0 | | 2.0 | 3.42 |
| gap_y | 0.0 | 5.0 | | 2.0 | 4.96 |
| top_x_shift_ant | 0.0 | 2.5 | | 0.5 | 1.48 |
| top_y_shift_lat | 0.0 | 1.5 | | 1.35 | 0.72 |
| top_x_shift_pos | 0.0 | 2.5 | | 2.0 | 0.85 |

## Motion-preservation zone

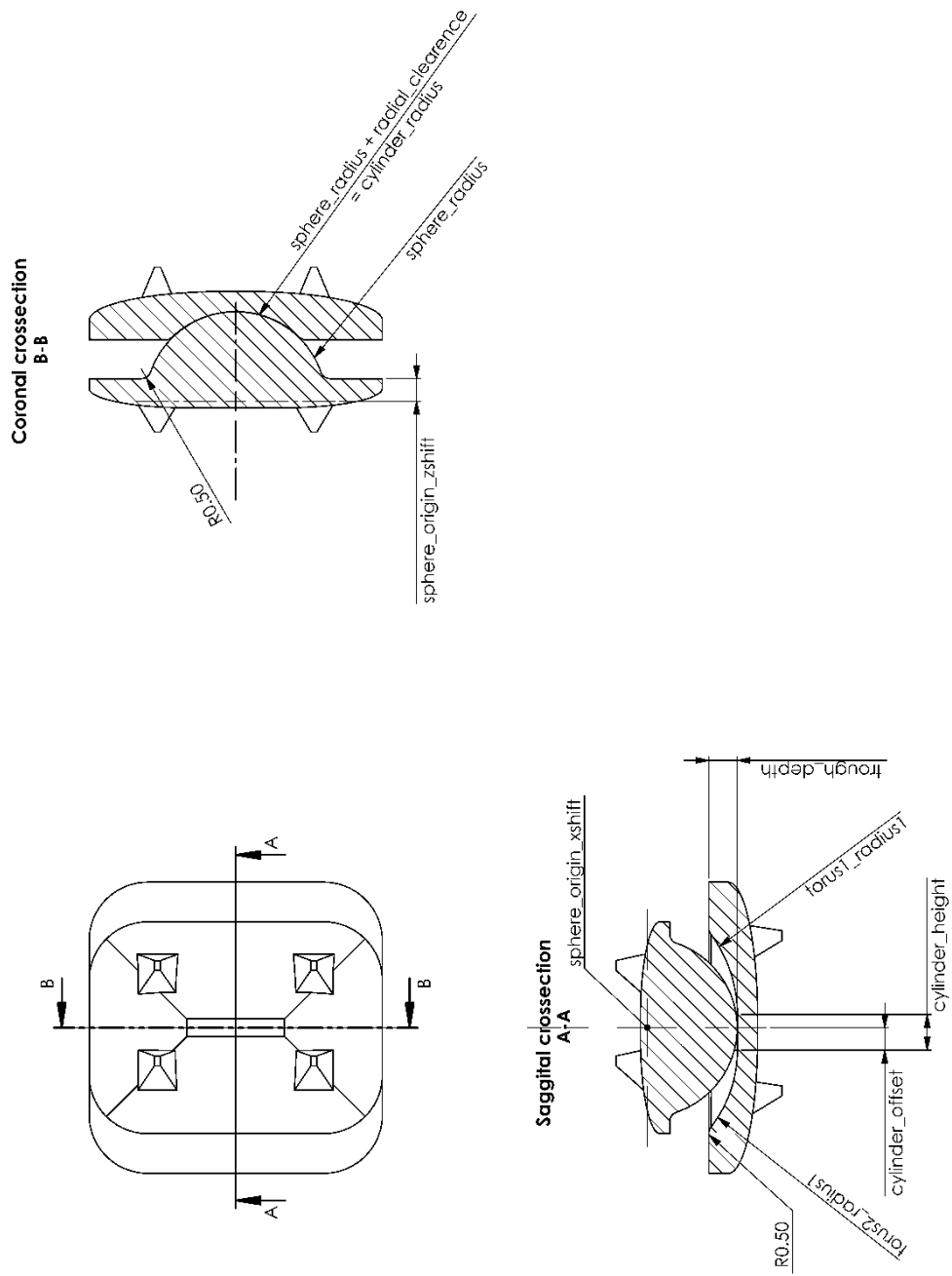

**Figure 9: The design variables of the single articulation motion-preservation zone optimization shown on the single articulation baseline design.**

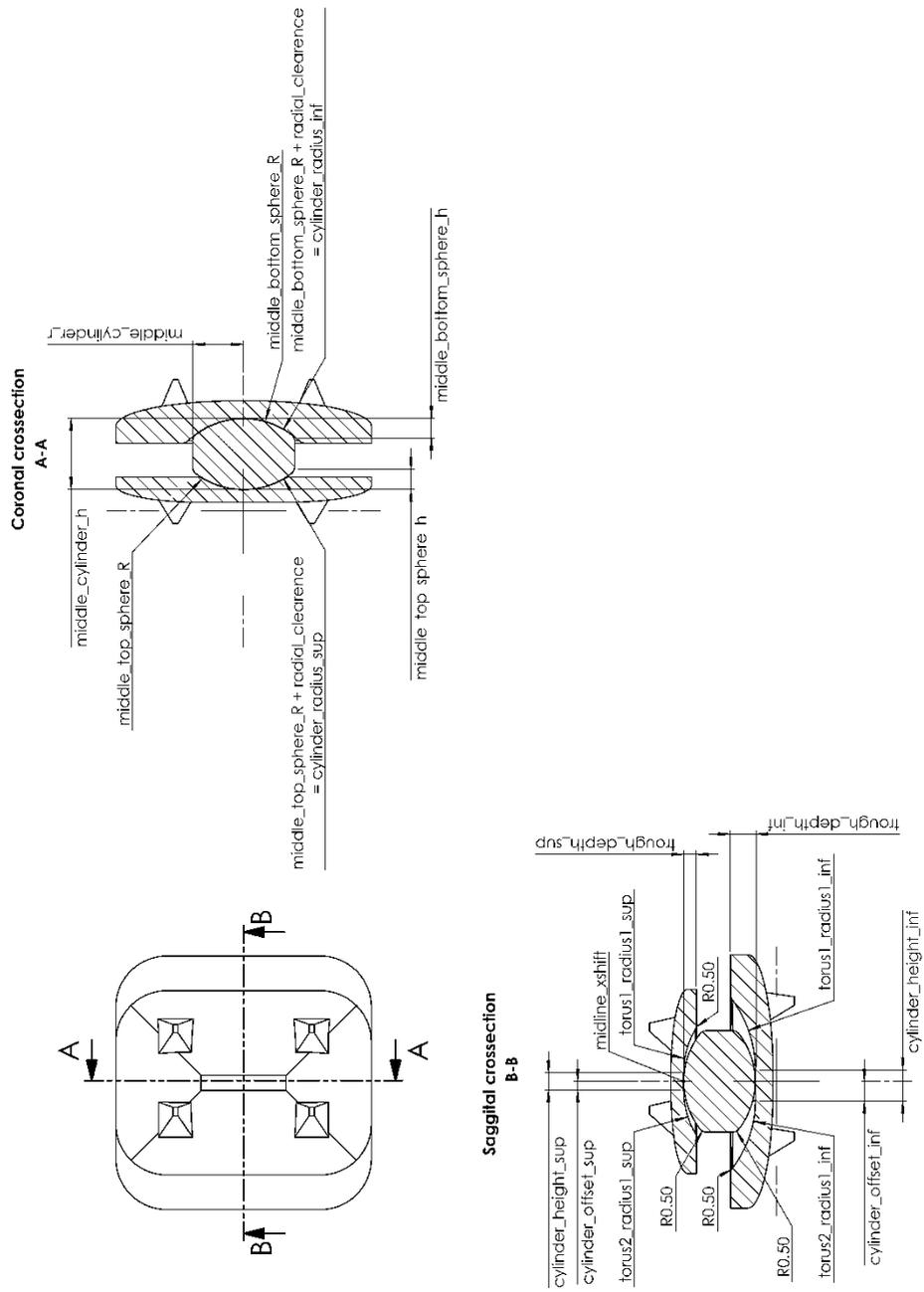

**Figure 10:** The design variables of the dual articulation motion-preservation zone optimization shown on the dual articulation baseline design.

**Table 2: Design space of the single articulating designs motion-preservation zone optimization.**

| Design variable | Min Value | Max Value | Baseline design | Optimized design |
|---|---|---|---|---|
| sphere_origin_xshift | -7.5 | 7.5 | 0.0 | -0.17 |
| sphere_origin_zshift | 0.0 | 5.85 | 1.16 | 2.39 |
| sphere_radius | 3.0 | 7.35 | 4.60 | 4.96 |
| cylinder_radius | dependent | | 4.67 | 5.03 |
| cylinder_height | 0.0 | 5.0 | 1.84 | 2.16 |
| cylinder_offset | -2.5 | 2.5 | -0.24 | 0.99 |
| trough_depth | dependent | | 1.45 | 0.58 |
| torus1_radius1 | 0.0 | 7.0 | 2.31 | 1.94 |
| torus2_radius1 | 0.0 | 7.0 | 1.82 | 0.02 |

**Table 3: Design space of the dual articulating designs motion-preservation zone optimization.**

| Design variable | Min Value | Max Value | Baseline design | Optimized design |
|---|---|---|---|---|
| midline_xshift | -3.95 | 3.95 | 0.0 | 0.21 |
| *Superior implant part: [...]_sup* | | | | |
| cylinder_radius | dependent | | 4.67 | 3.09 |
| cylinder_height | 0.0 | 5.0 | 1.0 | 0.58 |
| cylinder_offset | -2.5 | 2.5 | 0.0 | -0.27 |
| trough_depth | 0.5 | 1.0 | 0.75 | 0.84 |
| torus1_radius1 | 0.0 | 7.0 | 0.5 | 0.12 |
| torus2_radius1 | 0.0 | 7.0 | 0.5 | 2.04 |
| *Inferior implant part: [...]_inf* | | | | |
| cylinder_radius | dependent | | 4.67 | 3.10 |
| cylinder_height | 0.0 | 5.0 | 1.84 | 2.10 |
| cylinder_offset | -2.5 | 2.5 | -0.24 | 1.01 |
| trough_depth | 0.5 | 2.0 | 1.46 | 0.92 |
| torus1_radius1 | 0.0 | 7.0 | 2.31 | 6.73 |
| torus2_radius1 | 0.0 | 7.0 | 1.82 | 5.41 |
| *Mobile insert: middle_[...]* | | | | |
| cylinder_h | dependent | | 1.98 | 1.93 |
| cylinder_r | 1.0 | 4.95 | 3.0 | 2.17 |
| top_sphere_R | dependent | | 4.60 | 3.02 |
| top_sphere_h | 0.5 | 2.5 | 1.1 | 0.92 |
| bottom_sphere_R | dependent | | 4.60 | 3.03 |
| bottom_sphere_h | 0.5 | 2.5 | 1.1 | 0.92 |

## Appendix 2: Geometry creation

### Implant geometry

Figure 11 shows the geometry creation on the example of the baseline design that was used for the bone-implant interface optimization and single articulation motion-preservation zone optimization.

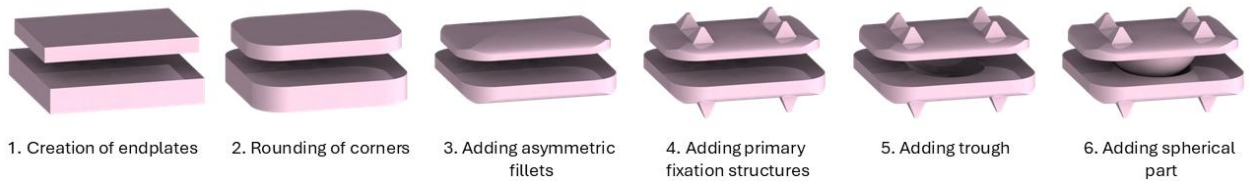

Figure 11: Geometry creation on the example of the baseline design of the bone-implant interface optimization and single articulation motion-preservation zone optimization. Fillets between the spherical part and the endplate and between the trough and the endplate aid manufacturing, but more importantly serve a mechanical purpose. These fillets are included in step 5 and 6.

### Vertebrae for the bone-implant interface optimization

Figure 12 shows the creation methodology of the vertebral body models used for the bone-implant interface optimization. To this end, we created the cortical and trabecular bone geometries from the preexisting VIVA+ bone geometries based on the thickness values in eight locations per bone (based on a radiographic study)[22] and created a geometry via a surface based on multiple planes.

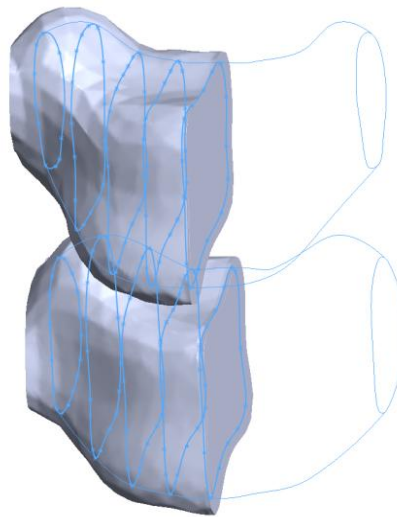

Figure 12: Creating the cortical and trabecular bone (separation approach shown in blue) from the vertebral body geometry (shown in a cross-section for visibility).

## Appendix 3: Load cases

The simulations started with a settling phase, during which a small initial gap between the implant parts was closed and all structures were moved into their initial position. Figure 13 and Figure 14 visualize the load cases used.

## Bone-implant interface

Figure 13 shows the subsidence and expulsion load cases used in the optimization of the bone-implant interface.

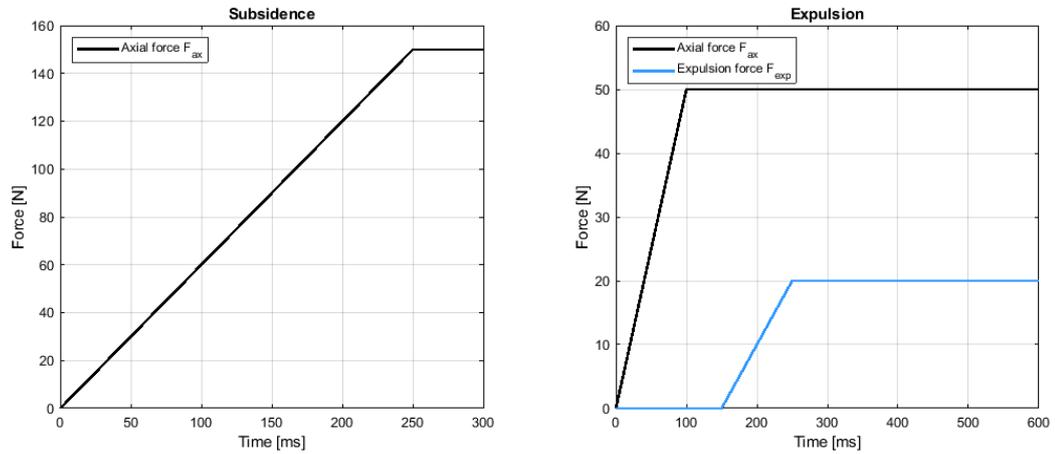

**Figure 13: Load case plots used in the bone-implant interface optimization. As illustrated in Figure 4, in the expulsion load case the expulsion force is applied to each of the two expulsion blocks.**

## Motion-preservation zone

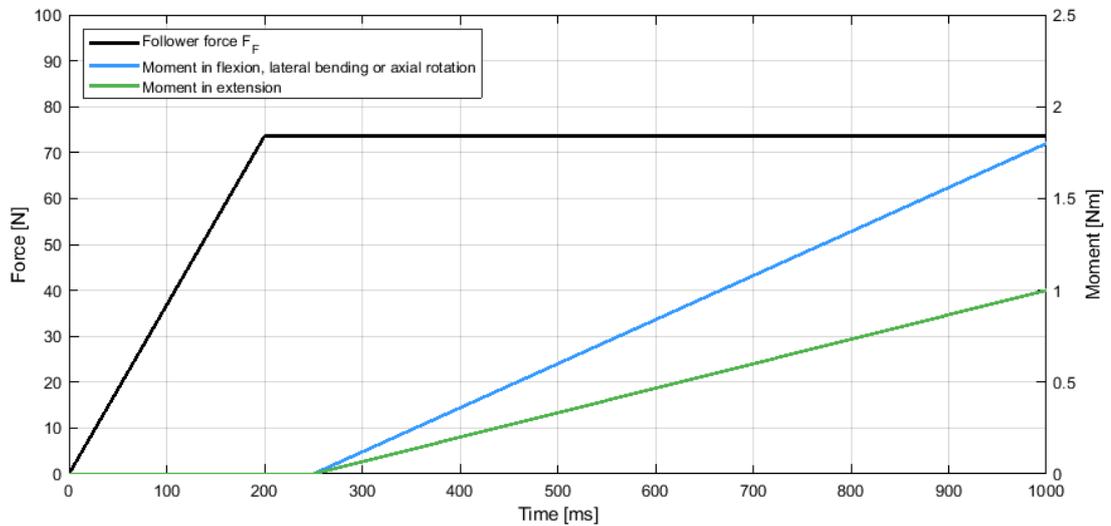

**Figure 14: Load case plots for the motion-preservation zone optimization.**

## Appendix 4: Termination criteria definitions

The design change criterion is defined as:

$$tol_p > \frac{\|p^{(k)} - p^{(k-1)}\|}{\|\Omega\|} \qquad (3a)$$

As defined before, the vector p contains the geometric design variables, which are subject to the design space Ω. The number of the iteration is given as k.

The objective function criterion is defined as:

$$tol_f > \left|\frac{f^{(k)} - f^{(k-1)}}{f^{(k-1)}}\right| \qquad (3b)$$

With f being the value of the objective function.

## Appendix 5: Typical subsidence and expulsion testing

Figure 15 shows typical subsidence and expulsion tests.

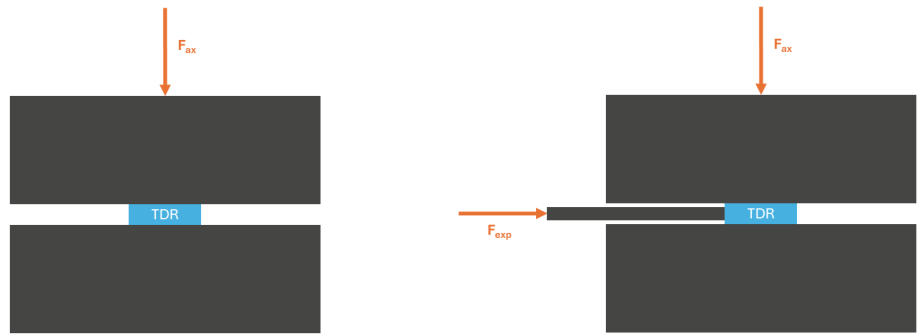

Figure 15: Schematic of typical subsidence (left) and expulsion (right) tests.

## Appendix 6: Comparison of the kinematics of the preoperative simulations to cadaveric and in *vivo* studies

An ex vivo study by Moroney et al.[33] using a preload of 73.6N reports 5.55 ± 1.84° of flexion and -1.54 ± 0.72mm posterior translation under 1.8Nm flexion moment, and 3.52 ± 1.94° of extension and 1.15 ± 1.01mm of posterior translation under 1.8Nm extension moment. They furthermore found 4.71 ± 2.99° of lateral bending under 1.8Nm lateral bending moment and 1.85 ± 0.67° of axial rotation under 1.8Nm axial rotation moment. During lateral bending, they measured 1.50 ± 1.24° of axial rotation, during axial rotation, they measured 0.95 ± 1.12° of lateral bending, see Table 4. While we apply a reduced moment in extension, we measured larger extension rotations than Moroney et al.

**Table 4: Kinematics of simulated motions.** The rows show the different motions that were simulated for the different conditions. The columns show the most relevant rotations and translations during these load cases in a local coordinate system.

| | | | Flexion / extension [°] | Lateral bending [°] | Axial Rotation [°] | Anterior/ posterior translation [mm] |
|---|---|---|---|---|---|---|
| Flexion | Preoperative | 73.6N; 1.8Nm | 5.48 | | | 1.03 |
| | Moroney et al. | 73.6N; 1.8Nm | 5.55 ± 1.84 | | | 1.54 ± 0.72 |
| Extension | Preoperative | 73.6N; 1.0Nm | -6.16 | | | -1.19 |
| | Moroney et al. | 73.6N; 1.8Nm | 3.52 ± 1.94 | | | 1.15 ± 1.01 |
| Lateral bending | Preoperative | 73.6N; 1.8Nm | | 3.20 | -1.73 | |
| | Moroney et al. | 73.6N; 1.8Nm | | 4.71 ± 2.99 | 1.50 ± 1.24 | |
| Axial rotation | Preoperative | 73.6N; 1.8Nm | | -1.95 | 2.49 | |
| | Moroney et al. | 73.6N; 1.8Nm | | 0.95 ± 1.12 | 1.85 ± 0.67 | |

A review on *in vivo* kinematics of the cervical spine reports maximal angular ranges of motion for C5/C6 to be 14.9° ± 5.4° (mean ± SD) in flexion/extension, 9.6° ±3.0° in lateral bending and 6.9° ± 2.6° in axial rotation[34]. Our preoperative values are within one SD of the mean in case of flexion/extension and axial rotation but slightly below it for lateral bending (0.2°). But our simulations were anyway not intended to reach the maximal ranges of motion but rather to be realistic of daily life. During most activities of daily living, only 20-40% of the maximal ranges of motion are used[35] While this is not specifically for C5/C6, we encompass that for all motions.

The previously mentioned review reports 1.9mm ± 0.9mm (mean ± SD) of coupled anteroposterior translation during flexion/extension motion[34]. The anteroposterior translation of the present study was 2.22mm which is within one standard deviation of the literature value.

# Appendix 7: Comparison of facet joint forces of the preoperative simulations to a cadaveric study

A cadaveric study (C2-T1) found facet joint forces of 27N in extension, 34,7N in left lateral bending and 32.8N in right axial rotation in the left facets of C5/C6 levels of intact specimens[36]. They do not report the facet joint forces for flexion. Comparability is limited as they tested multilevel specimens under 50N axial preload and ±2.0 Nm pure moment in flexion/extension, lateral bending and axial rotation and therefore the segmental ROM of C5/C6 in their study was 5.44° in flexion, 4.66° in extension, 4.68° in lateral bending and 7.16°s in axial rotation.

In the optimization, we record a sum of the left and right facet joint force. The preoperative model had 53.77N in extension, 47.41N in axial rotation and 8.04N in lateral bending. That we do not resect the facet joint capsule might cause a difference to experimental results where the capsule needs to be opened to insert a sensor. It makes sense that the sum of the forces is in some motions about twice that of a single side and in other motions about the same as a single side.